\documentclass[twocolumn,showpacs,showkeys,superscriptaddress]{revtex4}
\usepackage{graphicx}
\usepackage{bm}
\usepackage{color}
\usepackage{amsmath}
\usepackage{amsfonts}
\usepackage{amssymb}
\usepackage{natbib}
\usepackage{latexsym}
\usepackage{mathrsfs}

\begin{document}

\title{Light--like propagation of self--interacting Klein--Gordon fields in cosmology}

\author{Felipe A. Asenjo}
\email{felipe.asenjo@uai.cl}
\affiliation{Facultad de Ingenier\'ia y Ciencias,
Universidad Adolfo Ib\'a\~nez, Santiago 7491169, Chile.}
\author{Sergio A. Hojman}
\email{sergio.hojman@uai.cl}
\affiliation{Departamento de Ciencias, Facultad de Artes Liberales,
Universidad Adolfo Ib\'a\~nez, Santiago 7491169, Chile.}
\affiliation{Departamento de F\'{\i}sica, Facultad de Ciencias, Universidad de Chile,
Santiago 7800003, Chile.}
\affiliation{Centro de Recursos Educativos Avanzados,
CREA, Santiago 7500018, Chile.}

\begin{abstract}
It is showed that complex scalar fields with a self-interaction potential may propagate along null geodesics on flat Friedmann--Lema\^itre--Robertson--Walker universes with different time-dependent scale factors. This occurs provided they self interact adequately, for different forms of potentials, and 
 even for the massive case.
\end{abstract}

%\pacs{04.60 Bc, 98.80 Qc, 11.30 Cp}

\maketitle

Light-rays propagate in null geodesics, with four--wavevectors $k_\mu$ that satisfy the relation $k_\mu k^\mu=0$ on any curved spacetime background. Massless scalar fields share the same feature, having similar light-like four--wavevectors. However, are massless scalar fields the only ones that may propagate with light--like wavevectors? 

In a cosmological scenario, we approach this question by showing that scalar fields with specific self--interactions have null geodesic like propagation for different kinds of evolving cosmological isotropic flat Universes.
This behavior is exhibited by complex scalar fields $\phi$ propagating in the presence of self--interaction (real) potentials, given by 
\begin{equation}
    V=V\left(\phi^*\phi\right)=\lambda \left(\phi^*\phi\right)^n\, ,
    \label{potentialself}
\end{equation}
where $\lambda$ and $n$ are constants.

Consider an isotropic flat  cosmology, described by the  Friedmann--Lema\^itre--Robertson--Walker (FLRW) 
metric written in spherical coordinates
$ds^2=-dt^2+a^2 dr^2+a^2 \left(d\theta^2+\sin^2\theta d\phi^2\right)$, with a time--dependent scale factor $a=a(t)$. In this cosmological scenario, radial null geodesic motion ($ds^2=0$) is defined by 
\begin{equation}
    \frac{dr}{dt}=\frac{1}{a}\, .
    \label{geodlight}
\end{equation}
This equation may be used, for instance, to compute the  cosmological redshift of light  in a 
FLRW Universe \cite{ryder}.

The essence of this work is to determine what kinds of self--interactions allow complex
scalar fields to propagate with four--wavevectors that are characterized by null geodesic behavior ($k_\mu k^\mu=0$) satisfying Eq.~\eqref{geodlight} in a FLRW cosmology.
As we will show, scalar fields exhibit light--like propagation due to the non-local feature of fields that balances the self-interacting phenomena introduced by potential \eqref{potentialself}. The non-locality of the field is caused by the (non--vanishing) Bohm potential of the scalar field. For example, the non-local effects of Bohm potential in the propagation of fields have been shown (both theoretically and experimentally) to produce modifications to the null-geodesic propagation of light \cite{hojase1,hojase2,Mashhoon1,Mashhoon2,Mashhoon3,rivka} and scalar fields \cite{hojase5}, to the non-diffracting features of gravitational waves \cite{hojase3,hojase4},  and to quantum \cite{hojase6,hojase7,Makowski,electron} and optical \cite{inaoe,airy} behavior of different systems, among others.

Let us start with the dynamics of a complex scalar field $\phi$ in the FLRW background, following the equation $\Box \phi-{\partial V}/{\partial \phi^*}=0$,
where $\Box$ is the D'Alambertian operator in curved spacetime, and $V$ is the potential \eqref{potentialself}. The complex scalar field satisfies the equation
\begin{equation}
\Box \phi-\lambda n \left(\phi^*\phi\right)^{n-1}\phi=0\, .
\end{equation}
 A massless scalar field is obtained when $\lambda=0$. On the contrary,  the massive scalar field is recovered when $n=1$, and $\lambda=m^2$. This equation is not conformally invariant, in general. 
Now, we can write the scalar field using a polar representation as
\begin{equation}
    \phi=\varphi \exp\left(i S\right)\, ,
\end{equation}
where $\varphi$ represents a real time--dependent amplitude of the scalar field, and $S$ is its real phase. Then, the equations of motion for the field are now
\begin{eqnarray}
    k^\mu k_\mu+ \lambda n \varphi^{2n-2}&=&\frac{\Box\varphi}{\varphi}\label{ec1kkg}\, ,\\
    \nabla_\mu\left(k^\mu\varphi^2\right)&=&0\label{ec2kkg}\, ,
\end{eqnarray}
where the wavevector is given by $k_\mu=\nabla_\mu S$, where $\nabla_\mu$ is the covariant derivative. The right-hand side of Eq.~\eqref{ec1kkg} is the Bohm potential of the scalar field, which is the responsible for the scalar field non-locality. 

The above equations are general. Now, let us focus on a background given by an isotropic flat cosmology. We require that the above general complex scalar field propagates in a light--like fashion, i.e. along a null geodesic. This is achieved by requiring 
\begin{equation}
    k_\mu k^\mu=0\, .
\end{equation} 
Consider a field that propagates radially and assume that the phase $S$ depends on $r$ and $t$ only, with constant $k_r$. Then the above equation translates to $0=-k_0^2+k_r^2/{a^2}$, implying
\begin{equation}
\frac{1}{a}=\frac{k_0}{k_r}=\frac{\partial_0 S}{\partial_r S}=\frac{dr}{dt}\, .
\label{geodlight2}
\end{equation}
This is equivalent to the null geodesic propagation of light, given by Eq.~\eqref{geodlight}, which allows us to determine the phase as $S(t,r)=k_r r+k_r\int dt/a$.
Then, Eq.~\eqref{ec2kkg} can be used to find the amplitude of the scalar field, giving rise to
\begin{equation}
    \varphi(t)=\frac{\varphi_0}{a}\, .
\end{equation}
where $\varphi_0$ is an arbitrary constant.

In this way, we have been able to fully determine the general condition for light--like propagation of a complex scalar field in terms of the potential \eqref{potentialself} only. 
This is expressed through Eq.~\eqref{ec1kkg}, now written  as
\begin{eqnarray}
     \lambda n \varphi^{2n-2}&=&\frac{\Box\varphi}{\varphi}\, ,
\end{eqnarray}
where the effects of the self-interaction are balanced by the Bohm potential. This equation becomes in
\begin{eqnarray}
    \frac{n \lambda}{a^{2n-2}}=\frac{\ddot a}{a}+\frac{\dot a^2}{a^2}\, ,
    \label{ecprincipalricci}
\end{eqnarray}
where $\varphi_0$ has been absorbed in $\lambda$.
If solutions for $\lambda$ and $n$ can be found in Eq.~\eqref{ecprincipalricci} for different cosmologies, then the complex scalar field propagates along null geodesics in such cosmologies. 

The simplest form that we can consider for the time dependence of the  scale factor  is $a=t^\beta$, where  $\beta$ is a constant. This form for $a$ describe several kinds of different cosmological scenarios \cite{ryder}. Thereby, Eq.~\eqref{ecprincipalricci} becomes
${n \lambda}\, {t^{2+2\beta(1-n)}}={2\beta^2-\beta}$,
which is solved by
\begin{eqnarray}
    n&=&\frac{1}{\beta}+1\, ,\nonumber\\
    \lambda&=& \frac{\beta^2(2\beta-1)}{\beta+1}\, ,
    \label{ecprincipalricci3}
\end{eqnarray}
and thus, $\beta$ determines completely the self--interaction potential.

For a {\it radiation-dominated spatially flat universe} we have $a=t^{1/2}$ \cite{ryder}. Thus, $\beta=1/2$,  $\lambda=0$, and the potential is
\begin{equation}
    V=0\, .
\end{equation}
Consequently, in this cosmological scenario, only a massless complex scalar field can propagate in null geodesics.

In a more  general scenario, we may study a 
{\it general spatially flat universe}, where the scale factor evolves as
$\beta={2}/({3+3w})$ \cite{ryder},
where $w=P/\varepsilon$ determines the equation of state  of a cosmological fluid with pressure $P$ and energy density $\varepsilon$ ($-1< w\leq 1$). In this way,
from Eq.~\eqref{ecprincipalricci3} we obtain  
 \begin{eqnarray}
 n&=&\frac{3w+5}{2}\, ,\nonumber\\
     \lambda&=&\frac{4(1-3w)}{9(1+w)^2(5+3w)}\, .
     \label{lambdayn}
 \end{eqnarray}
Notice that  in solution \eqref{lambdayn}, $\lambda>0$ for $w<1/3$. For other $w>1/3$, $\lambda$ is negative.

For instance, for a {\it matter-dominated universe}, with $w=0$, a complex scalar field propagates in this cosmology along a null geodesic in the presence of the self-interaction potential
\begin{equation}
    V=\frac{4}{45}\left(\phi^*\phi\right)^{5/2}\, .
\end{equation}
On the other hand, for a negative energy density fluid, with the equation of state $w=-1/3$, a complex scalar field propagates in this cosmology if it has the potential
\begin{equation}
    V=\frac{1}{2}\left(\phi^*\phi\right)^{2}\, .
\end{equation}
 Finally, for the case with $w=1$ the potential acquires the form
\begin{equation}
    V=-\frac{1}{36}\left(\phi^*\phi\right)^{4}\, .
\end{equation}

On the other hand, the dark energy cosmological case, with $w=-1$, deserves a separate study. 
In this case, this
{\it exponentially expanding spatially flat universe} has a scale factor $a=\exp\left(H t\right)$, with a  Hubble constant $H$ \cite{ryder}. Using this in Eq.~\eqref{ecprincipalricci}, we find that $n=1$ and $\lambda=2 H^2$.
Therefore, the scalar field self--interacts through the potential
\begin{equation}
    V=2 H^2 \left(\phi^*\phi\right)\, .
\end{equation}
This implies that a massive complex scalar field, with a mass with the specific value of
\begin{equation}
    m=\sqrt{2} H\, .
\end{equation}
may propagate along null geodesics in a dark energy cosmology.
This result was first found by Molski \cite{molski}, establishing that
a massive particle can indeed propagate in a light-like fashion in a cosmological context.

%%%%%%%

The above results
show that any complex scalar field, with a proper self--interaction potential, can follow null geodesics in different cosmological scenarios. This is due to the balance of the effects of a self-interaction potential and the non--local features of the field as described by the non--vanishing Bohm potential.
It is that non-locality which forces the scalar field to have the light--like propagation, with $k_\mu k^\mu=0$.

Along this work we have used a potential with the form 
\eqref{potentialself}. However
different forms of
self-interaction may be considered to produce these kinds of behavior \cite{hojase3}. Besides, any field is non-local and, consequently, this form of null geodesic behavior may be present for other self--interacting fields in curved spacetime backgrounds.


\begin{thebibliography}{}

\bibitem{ryder} B. Ryden, {\it Introduction to Cosmology} (Addison Wesley, 2003).

\bibitem{hojase1} F. A. Asenjo and S. A. Hojman, Class. Quantum Grav. {\bf 34}, 205011  (2017).
\bibitem{hojase2} F. A. Asenjo and S. A. Hojman, Phys. Rev. D {\bf 96}, 044021 (2017).
\bibitem{Mashhoon1} B. Mashhoon, Phys. Rev. D {\bf 7}, 2807 (1973).
\bibitem{Mashhoon2} B. Mashhoon, Phys. Rev. D {\bf 11}, 2679 (1975).
\bibitem{Mashhoon3} B. Mashhoon, Phys. Lett. A {\bf 122}, 299 (1986).
\bibitem{rivka} A. Patsyk, M. A. Bandres, R. Bekenstein and M. Segev, Phys. Rev. X {\bf 8}, 011001 (2018).
\bibitem{hojase5} F. A. Asenjo and S. A. Hojman, Eur. Phys. J. C {\bf 81}, 98 (2021).
\bibitem{hojase3} F. A. Asenjo and S. A. Hojman, Eur. Phys. J. C {\bf 77}, 732 (2017).
\bibitem{hojase4} S. A. Hojman and F. A. Asenjo, Phys. Scr. {\bf 95}, 085001  (2020).
\bibitem{hojase6} S. A. Hojman and F. A. Asenjo, Phys. Lett. A {\bf 384}, 126913
(2020).
\bibitem{hojase7} S. A. Hojman and F. A. Asenjo, Phys. Rev. A, {\bf 102}, 052211 (2020).
\bibitem{Makowski} A. J. Makowski and S. Konkel, Phys. Rev. A {\bf 58}, 4975 (1998).

\bibitem{electron} N. Voloch-Bloch, Y. Lereah, Y. Lilach, A. Gover and A. Arie, Nature {\bf 494}, 331 (2013).
\bibitem{inaoe} F. A. Asenjo, S. A. Hojman, H. M. Moya-Cessa and F. Soto-Eguibar, 
Opt. Comm. {\bf 490}, 126947  (2021).

\bibitem{airy} G. A. Siviloglou, J. Broky, A. Dogariu and D. N. Christodoulides,
Phys. Rev. Lett. {\bf 99}, 213901 (2007).






\bibitem{molski} M. Molski, Phys. Essays {\bf 7}, 99 (1994).



\end{thebibliography}
\end{document}